\documentstyle[preprint,aps,epsfig]{revtex}

\tightenlines
\preprint{\vbox{  
\hbox{IFT-P.096/2000}   
\hbox{June 2001} }
}  
\begin{document}  
\draft 
\title{
A supersymmetric 3-3-1 model}
\author{J. C. Montero, V. Pleitez and M. C. Rodriguez} 
\address{
Instituto de F\'\i sica  Te\'orica\\ 
Universidade  Estadual Paulista\\
Rua Pamplona, 145\\ 
01405-900-- S\~ao Paulo, SP\\ 
Brazil } 
\maketitle 

\begin{abstract}  
We build the complete supersymmetric version of a 3-3-1 gauge model
using the superfield formalism. 
We point out that a discrete symmetry, similar to the R-symmetry in the 
minimal supersymmetric standard model, is possible to be defined in this 
model. Hence we have both R-conserving and R-violating possibilities.
Analysis of the mass spectrum of the neutral real scalar fields
shown that in this model the lightest scalar Higgs has a mass upper limit, and 
at the tree level it is 124.5 GeV for a given illustrative set of parameters.  
\end{abstract}

\section{Introduction}
\label{sec:intro}

Although the standard model (SM), based on the gauge symmetry 
$SU(3)_c\otimes SU(2)_L\otimes U(1)_Y$ describes the observed properties
of charged leptons and quarks it is not the ultimate theory. 
However, the necessity to go beyond it, from the 
experimental point of view, comes at the moment only from neutrino 
data~\cite{bk}. 
If neutrinos are massive then new physics beyond the SM is needed.
From the theoretical point of view, the SM cannot be a fundamental theory 
since it has so many parameters and some important questions like that of 
the number of families do not have an answer in its context.
On the other side, it is not clear what the physics beyond the SM should be.
An interesting possibility is that at the TeV scale physics would be 
described by models which share some of the faults of the SM but give
some insight concerning some questions which remain open in the SM context.

One of these  possibilities is that, at energies of a few TeVs, the gauge
symmetry may be $ SU(3)_c\otimes SU(3)_L \otimes U(1)_N$ (3-3-1 for shortness)
instead of that of the SM~\cite{pp,pf}. In fact, this may be the
last symmetry involving the lightest elementary particles: leptons. 
The lepton sector is exactly the same as in the SM but now there is a 
symmetry, at large energies among, say 
$e^-$, $\nu_e$ and $e^+$. Once this
symmetry is imposed on the lightest generation and extended to the other 
leptonic generations it follows that the quark sector must be enlarged 
by considering exotic charged quarks.  It
means that some gauge bosons carry lepton and baryon quantum number.
Although this model coincides at low energies with the SM it explains 
some fundamental questions that are accommodated, but not explained, in the 
SM. These questions are:

{\it i)} The family number must be a multiple of three in order to cancel 
anomalies~\cite{pp,pf}.
This result comes from the fact that the model is anomaly-free only if we 
have equal number of triplets and antitriplets, counting the 
$SU(3)_{c}$ colors, and further more requiring the sum of all fermion charges 
to vanish. However each generation is anomalous, the anomaly 
cancellation occurs for the three, or multiply of three, together and not
generation by generation like in the SM. This may
provides a first step towards answering the flavor question. 

{\it ii)} Why $\sin^{2} \theta_{W}<\frac{1}{4}$ is observed.
This point come from the fact that in the model of Ref.~\cite{pp} we have that
the $U(1)_N$ and $SU(3)_L$ coupling constants, $g'$ and $g$, respectively,
are related by  
\begin{equation}
t^{2}\equiv\frac{g^{\prime 2}}{g^2}= 
\frac{ \sin^{2} \theta_{W}}{1-4 \sin^{2} \theta_{W}}.
\label{eq2}
\end{equation}
Hence, this 3-3-1 model predicts that there exists an energy scale, 
say $\mu$, at which the model loses its perturbative character. 
The value of $\mu$ can be found through the condition   
$\sin^2\theta_W(\mu)=1/4$. However, it is not clear at all 
what is the value of $\mu$; in fact, it has been argued that 
the upper limit on the vector bilepton masses is 3.5 TeV~\cite{jj} 
instead of the 600 GeV given in Ref.~\cite{phf}. 
Any way, the important point is that in this model the ``hierarchy 
problem'' i.e., the existence of quite different mass scales, is less 
severe than in the SM, and its extensions, since now
no arbitrary mass scale can be introduced in the
model. Hence, in
this supersymmetric 3-3-1 model (thereafter called 3-3-1s for short) it is
natural that supersymmetry is broken at the TeV scale 
This is very important because one of the motivation for supersymmetry
is that it can help to understand the hierarchy problem: if it is
broken at the TeV scale. Notwithstanding, in the context
of the SM it is necessary to assume that the breakdown of
supersymmetry happens at the TeV scale. 
However, other 3-3-1 models i,e., with different representation content, have
different upper limit for the maximal energy scale~\cite{mpp}. 

{\it iii)} The quantization of the electric charge is
possible even in the SM context. This
is because of the classical (hypercharge invariance of the 
Yukawa interactions) and quantum constraints (anomalies)~\cite{cq1}.
However this occurs only family by family and if there is no  
right-handed neutrinos (neutrinos if massive must be Majorana fields); 
or, when the three families are 
considered together the quantization of the electric charge is possible only
if right-handed neutrinos with Majorana mass term are introduced~\cite{rf}
or another Higgs doublet~\cite{2h} or some neutral fermions~\cite{nf} are
introduced. On the other hand, in the 3-3-1 model~\cite{pp,pf,mpp} the charge
quantization in the three families case does not depend if neutrinos are 
massless or massive particles~\cite{cp1}.

{\it iv)} In the context of the SM with only one generation, as in the 
previous item, both classical and quantum constraints imply that  
the quantization of the charge and the vectorial nature of the 
electromagnetic charge arise together.
When right-handed neutrinos are added there is no charge quantization but 
the vectorial nature of electromagnetic interactions survives.
Both of them are restored if neutrinos are Majorana particles~\cite{cq1}.
In the three generation case neutrinos ought to be also Majorana particles
in order to retain both features of the electromagnetic 
interactions~\cite{rf,cq2}. 
On the other hand, in all sort of 3-3-1 models the quantization of 
the charge and the vectorial nature of the electromagnetic interactions 
are related one to another and are also independent of the nature of the
neutrinos~\cite{cp2}. 

Last but not least, {\it v)} if we accept the criterion that particle 
symmetries are 
determined by the known leptonic sector, and if each generation is treated 
separately, then $SU(3)_L$ 
is the largest chiral symmetry group to be considered among $(\nu,e,e^c)_L$.
The lepton family quantum number is gauged; only the total lepton number, $L$,
remains a global quantum number (or equivalently we can define ${\cal F}=
B+L$ as the global conserved quantum number where $B$ is the 
baryonic number~\cite{pt1}). On the other hand, if right-handed neutrinos 
do exist, as it appears to be the case~\cite{bk}, the symmetry among 
$(\nu_e,e,\nu^c_e,e^c)_L$ would be $SU(4)_L\otimes U(1)_N$~\cite{pp2}. 
This is possibly the last symmetry among leptons. There is no room for 
$SU(5)_L\otimes U(1)_N$ if we restrict ourselves to the case of leptons with 
electric charges $\pm1,0$~\cite{vp}. Hence, in this case all versions of the
3-3-1 model, for instance the one in Refs.~\cite{pp,pf,pt2}, and the one in 
Refs.~ \cite{mpp,valle,flt}, are different $SU(3)_L$-projections of the larger 
$SU(4)_L$ symmetry~\cite{pp2}.   

Besides the characteristic features given above, which we can consider 
predictions of the model, the model has some interesting phenomenological 
consequences:
a) An extended version of some 3-3-1 models solve the strong 
CP problem.
It was shown by Pal~\cite{pal1} that in 3-3-1 models~\cite{pp,pf,mpp,flt} the 
more general Yukawa couplings admit a Peccei-Quinn symmetry~\cite{pq} 
and that symmetry can be extended to the Higgs potential and, 
therefore, making it 
a symmetry of the entire lagrangian. This is obtained by introducing
extra Higgs scalar multiplets transforming under the 3-3-1 symmetry as 
$\Delta\sim({\bf1},{\bf10},-3)$, for the model of Refs.~\cite{pp,pf},
or $\Delta\sim({\bf1},{\bf10},-1)$, for the model of Refs.~\cite{mpp,flt}. 
In this case the resulting axion can be
made invisible. The interesting thing is that in those sort of models the 
Peccei-Quinn symmetry is an automatic symmetry, in the sense that it does 
not have to be imposed separately on the lagrangian but it is a consequence 
of the gauge symmetry and a discrete symmetry. 
b) There exist new contributions to the neutrinoless double beta
decay in models with three scalar triplets~\cite{pt1} or in the model
with the sextet~\cite{bbd331}. If the model is extended with a neutral
scalar singlet it is possible to have a safe Majoron-like Goldstone boson and
there are also contributions to that decay with Majoron emission~\cite{bbd331}.
c) It is the simplest model that includes bileptons of both types: 
scalar and vector ones.
In fact, although there are several models which include doubly charged
scalar fields, not many of them incorporate doubly
charged vector bosons: this is a particularity of the 3-3-1 model
of Refs.~\cite{pp,pf}.
d) The model has several sources of CP violation.
In the 3-3-1 model~\cite{pp,pf} we can implement the violation of the CP
symmetry, spontaneously~\cite{laplata1,dumm1} or explicitly~\cite{liung}. 
In models with exotic leptons it is possible to implement soft CP 
violation~\cite{cp3}. e) The extra neutral vector boson $Z^\prime$ conserves
flavor in the leptonic but not in the quark sector. The couplings to the leptons
are leptophobic because of the suppression factor $(1-4s^2_W)^{1/2}$ but with 
some quarks there are enhancements because of the factor $(1-4s^2_W)^{-1/2}$
~\cite{dumm2}.
f) Although the minimal scalar sector of the model is rather 
complicated, with at least three triplets, we would like to stress that 
it contains all extension of the electroweak 
standard model with extra scalar fields: two or more doublets~\cite{hhunter}, 
neutral gauge singlet~\cite{cmp}, or doubly charged scalar fields~\cite{zee},
or a combination of all that. However, some couplings which are allowed in the
multi-Higgs extensions are not in the present model when we consider an
$SU(2)\otimes U(1)$ subgroup. Inversely, there are some interactions that are
allowed in the present models that are not in the multi-Higgs extensions of the
SM, for instance, trilinear couplings among the doublets which have
no analog in the SM, or even the in MSSM. It means that the model preserves 
the memory of the 3-3-1 original symmetry.
Hence, in our opinion, the large Higgs sector is not an intrinsic trouble of
this model. 
g) Even if we restrict ourselves to leptons of charge $0,\pm1$ 
we can have exotic neutral~\cite{valle} or charged heavy leptons~\cite{pt2}.
h) Neutrinos can gain Majorana masses if we allow one of the 
neutral components of the scalar sextet to gain a non-zero vacuum expectation
value~\cite{ma1}, or if we introduce right-handed neutrinos~\cite{bbd800},
or if we add either terms in the scalar potential that break the total lepton 
number or an extra charged lepton transforming as singlet under the
3-3-1 symmetry~\cite{numass}.  

Of course, some of the goodness of this type of models, like that in items
{\it i}), {\it ii}) and {\it iv}) above, can be considered only as a hint to 
the final resolution of those
problems: they depend on the representation content and we can always ask 
ourselves what is the main principle behind the representation content.  
Anyway, we think that the 3-3-1 models have interesting features by 
themselves and that it is well motivated to generalize them by 
introducing supersymmetry. In the present paper we built exhaustively
the supersymmetric version of the 3-3-1 model of Refs.~\cite{pp,pf}.

The outline of the paper is as follows. In Sec.~\ref{sec:sm} we present the
representation content of the supersymmetric 3-3-1 model. 
We build the lagrangian in Sec.~\ref{sec:lagrangian}. In Sec.~\ref{sec:sp} we
analyze the scalar potential, in particular, we found the mass spectrum of the 
neutral scalar and shown that the lightest scalar field has an upper
limit of 124.5 GeV. Our conclusion are in the last section.

\section{The supersymmetric model}
\label{sec:sm}

The fact that in the 3-3-1 model of Refs.~\cite{pp,pf} we have the constraint
at tree  level $\sin^2\theta_W<1/4$ means, as we said before, that the
model predicts that there exists an energy scale, 
say $\mu$, at which the model loses its perturbative character.
Thus in that model the ``hierarchy 
problem'' i.e., the existence of quite different mass scales, is less 
severe than in the standard model and its extensions since
no arbitrary mass scale can be introduced in the
model. 
This feature remains valid when we introduce supersymmetry in the 
model. Thus, the breaking of the supersymmetry occurs 
in a natural way also at the TeV scale in this 3-3-1 model. 
Some aspects of the supersymmetric 3-3-1 model
have been already considered in Refs.~\cite{ema1,pal2} and we will comment on
later. 

However, let us first consider the particle content
of the model without supersymmetry~\cite{pp,pf}. We have the leptons
transforming as 
\begin{eqnarray}
   L_{l} &=& 
      \left( \begin{array}{c} \nu_{l} \\ 
                  l \\
                  l^{c}          
\end{array} \right)_{L} \sim ({\bf1},{\bf3},0), \,\ l= e, \mu , \tau.
\label{trip}
\end{eqnarray}
In parenthesis it appears the transformations properties under the respective
factors $(SU(3)_C,SU(3)_L,U(1)_N)$.
We have not introduced right-handed
neutrinos and for the moment we assume here that the
neutrinos are massless, however see~\cite{ma1,bbd800,numass}. 

In the quark sector, one quark family is also put in the triplet 
representation 
\begin{eqnarray}
   Q_{1L} &=& 
      \left( \begin{array}{c} u_{1} \\ 
                  d_{1} \\
                  J          
\end{array} \right)_{L} \sim \left({\bf3},{\bf3},\frac{2}{3}\right) \,\ , 
\label{q1l}
\end{eqnarray}
and the respective singlets are given by
\begin{equation}
u^{c}_{1L} \sim \left({\bf3}^*,{\bf1},-\frac{2}{3}\right),\quad
d^{c}_{1L} \sim \left({\bf3}^*,{\bf1},\frac{1}{3}\right),\quad  
J^{c}_{L} \sim \left({\bf3}^*,{\bf1},-\frac{5}{3}\right),
\label{q1r}
\end{equation}
writing all the fields as left-handed.

The others two quark generations we put in the antitriplet 
representation
\begin{equation}
\begin{array}{cc}  
Q_{2L} = 
      \left( \begin{array}{c} d_{2} \\ 
                  u_{2} \\
                  j_{1}          \end{array} \right)_{L},\quad  
 
  Q_{3L} = 
      \left( \begin{array}{c} d_{3} \\ 
                  u_{3} \\
                  j_{2}          \end{array} \right)_{L} 
\sim \left({\bf3},{\bf3}^{*},-\frac{1}{3}\right) , 
\end{array}
\label{q23l}
\end{equation}
and also with the respective singlets,
\begin{equation}
u^{c}_{2L} \,\ , u^{c}_{3L} \sim \left({\bf3}^*,{\bf1},-\frac{2}{3}\right),\quad
d^{c}_{2L} \,\ , d^{c}_{3L} \sim \left({\bf3}^*,{\bf1},\frac{1}{3}\right), \quad
j^{c}_{1L} \,\ , j^{c}_{2L} \,\ \sim \left({\bf3}^*,{\bf1},\frac{4}{3}
\right) \,\ . 
\label{q23r}
\end{equation}

On the other hand, the scalars which are necessary to generate the 
fermion masses are
\begin{equation} 
\eta = 
      \left( \begin{array}{c} \eta^{0} \\ 
                  \eta^{-}_{1} \\
                  \eta^{+}_{2}          \end{array} \right) \sim 
({\bf1},{\bf3},0),\quad 
\rho = 
      \left( \begin{array}{c} \rho^{+} \\ 
                  \rho^{0} \\
                  \rho^{++}          \end{array} \right) \sim 
({\bf1},{\bf3},+1),\quad 
\chi = 
      \left( \begin{array}{c} \chi^{-} \\ 
                  \chi^{--} \\
                  \chi^{0}          \end{array} \right) \sim 
({\bf1},{\bf3},-1),
\label{3t} 
\end{equation}
and one way to obtain an arbitrary mass matrix for the leptons is to introduce 
the following symmetric anti-sextet
\begin{equation}
S = 
      \left( \begin{array}{ccc} 
\sigma^{0}_{1}& \frac{h^{+}_{2}}{ \sqrt{2}}& \frac{h^{-}_{1}}{ \sqrt{2}} \\ 
\frac{h^{+}_{2}}{ \sqrt{2}}& H^{++}_{1}& \frac{ \sigma^{0}_{2}}{ \sqrt{2}} \\
 \frac{h^{-}_{1}}{ \sqrt{2}}& 
\frac{\sigma^{0}_{2}}{ \sqrt{2}}&  H^{--}_{2}        
\end{array} \right) \sim ({\bf1},{\bf6}^{*},0) \,\ . 
\label{sextet}
\end{equation}

Now, we introduce the minimal set of particles in order to implement the
supersymmetry~\cite{haber}. We have the sleptons corresponding to the leptons in
Eq.~(\ref{trip}); squarks related to the quarks in Eqs.(\ref{q1r})-(\ref{q23r});
and the Higgsinos related to the scalars given in Eqs.~(\ref{3t}) and
(\ref{sextet}). Besides, in order to to cancel chiral anomalies generated by the
superpartners of the scalars, we have to add the following higgsinos in the
respective anti-multiplets, 
\begin{mathletters}
\label{escac}
\begin{equation} 
\tilde{\eta}^{\prime} = 
      \left( \begin{array}{c} \tilde{\eta}^{\prime0} \\ 
                  \tilde{\eta}^{\prime+}_{1} \\
                  \tilde{\eta}^{\prime-}_{2}          \end{array} \right)_L 
\sim ({\bf1},{\bf3}^{*},0),\quad
\tilde{\rho}^{\prime} = 
      \left( \begin{array}{c} \tilde{\rho}^{\prime-} \\ 
                  \tilde{\rho}^{\prime0} \\
                  \tilde{\rho}^{\prime--}          \end{array} \right)_L 
\sim ({\bf1},{\bf3}^{*},-1),\quad 
\tilde{\chi}^{\prime} = 
      \left( \begin{array}{c} \tilde{\chi}^{\prime+} \\ 
                  \tilde{\chi}^{\prime++} \\
                  \tilde{\chi}^{\prime0}          \end{array} \right)_L 
\sim ({\bf1},{\bf3}^{*},+1),
\label{shtc}  
\end{equation}
\begin{equation}
\tilde{S}^{\prime} = 
      \left( \begin{array}{ccc} 
\tilde{\sigma}^{\prime0}_{1}& \frac{\tilde{h}^{\prime-}_{2}}{ \sqrt{2}}& 
\frac{\tilde{h}^{\prime+}_{1}}{ \sqrt{2}} \\ 
\frac{\tilde{h}^{\prime-}_{2}}{ \sqrt{2}}& \tilde{H}^{\prime--}_{1}& 
\frac{ \tilde{\sigma}^{ \prime 0}_{2}}{ \sqrt{2}} \\
 \frac{\tilde{h}^{\prime+}_{1}}{ \sqrt{2}}& 
\frac{ \tilde{\sigma}^{ \prime 0}_{2}}{ \sqrt{2}}&  
\tilde{H}^{\prime++}_{2}        
\end{array} \right)_L \sim ({\bf1},{\bf6},0).
\label{shsc} 
\end{equation}
\end{mathletters}

There are also the scalar partners of the
Higgsinos defined in Eq.~(\ref{escac}) and we will denote them $\eta^\prime,
\rho^\prime, \chi^\prime,S^\prime$. This is the particle
content which we will consider as the minimal 3-3-1s model if the charged lepton
masses are generated by the sextet $S$. 

Summaryzing, we have in the 3-3-1 supersymmetric model the following 
superfields:
$\hat{L}_{e,\mu,\tau}$, $\hat{Q}_{1,2,3}$, $\hat{\eta}$, $\hat{\rho}$, 
$\hat{\chi}$, $\hat{S}$; $\hat{\eta}^\prime$, $\hat{\rho}^\prime$, 
$\hat{\chi}^\prime$, $\hat{S}^\prime$;
$\hat{u}^c_{1,2,3}$, 
$\hat{d}^c_{1,2,3}$, $\hat{J}$ and $\hat{j}_{1,2}$, i.e., 23 chiral
superfields, and 17 vector superfields: $\hat{V}^a$, $\hat{V}^\alpha$ and
$\hat{V}^\prime$. In the minimal supersymmetric standard model (MSSM) there are
14 chiral superfields and 12 vector superfields.

\section{The Lagrangian}
\label{sec:lagrangian}

With the superfields introduced in the last section we can built a 
supersymmetric invariant lagrangian. It has the following form
\begin{equation}
   {\cal L}_{331} = {\cal L}_{SUSY} + {\cal L}_{soft}.
\label{l1}
\end{equation}
Here ${\cal L}_{SUSY}$ is the supersymmetric piece, while ${\cal L}_{soft}$ 
explicitly breaks SUSY.
Below we will write each of these lagrangians in terms of the 
respective superfields. 

\subsection{The Supersymmetric Term.}
\label{subsec:st}

The supersymmetric term can be divided as follows
\begin{equation} 
   {\cal L}_{SUSY} =   {\cal L}_{Lepton}
                  + {\cal L}_{Quarks} 
                  + {\cal L}_{Gauge} 
                  + {\cal L}_{Scalar}, 
\label{l2}
\end{equation}

where each term is given by

 \begin{equation} 
  {\cal L}_{Lepton} 
     = \int d^{4}\theta\;\left[\,\hat{ \bar{L}}e^{2g\hat{V}} 
 \hat{L} \,\right], 
\label{l3}
\end{equation}

\begin{eqnarray}
{\cal L}_{Quarks} 
     &=& \int d^{4}\theta\;\left[\,\hat{ \bar{Q}}_{1}
e^{2[g(\hat{V}_{c}+\hat{V})+(2g'/3)\hat{V}']} \hat{Q}_{1} 
+\,\hat{\bar{Q}}_{\alpha}
e^{2[g(\hat{V}_{c}+\hat{V})-(g'/3)\hat{V}']} \hat{Q}_{\alpha} \,\right. 
\nonumber \\&+& 
\left.\,\hat{ \bar{u}}_{i}
e^{2[g(\hat{V}_{c}+\hat{V})-(2g'/3)\hat{V}']} \hat{u}_{i} 
+\hat{ \bar{d}}_{i}
e^{2[g(\hat{V}_{c}+\hat{V})+(g'/3)\hat{V}']} \hat{d}_{i}\right.
\nonumber \\ &+&\left. \,\hat{ \bar{J}}
e^{2[g(\hat{V}_{c}+\hat{V}) -(5g'/3)\hat{V}']} \hat{J} 
+ \hat{ \bar{j}}_i
e^{2[g(\hat{V}_{c}+\hat{V}) +(4g'/3)\hat{V}']} \hat{j}_i\right]
\label{l4}
\end{eqnarray}

and

\begin{equation} 
  {\cal L}_{Gauge} = \frac{1}{4} \left[ \int  d^{2}\theta\; 
         [ {W}^{a}_{c}{W}^{a}_{c}+{W}^{a}{W}^{a}+{W}^{ \prime}{W}^{ \prime}
	   \int  d^{2}\bar{\theta}\; 
         [ \bar{W}^{a}_{c}\bar{W}^{a}_{c}+\bar{W}^{a}\bar{W}^{a}+
 \bar{W}^{ \prime}\bar{W}^{ \prime}]\,\right],
\label{l5}
\end{equation}
where $\hat{V}_{c}=T^{a}\hat{V}^{a}_{c}$, 
$\hat{V}=T^{a}\hat{V}^{a}$ and $T^a=\lambda^{a}/2$ are the generators of 
$SU(3)$ i.e., $a=1,\cdots,8$, and $g$ and $g^{\prime}$ are the gauge 
coupling of $SU(3)_L$ and $U(1)_N$. $W^{a}_{c}$, $W^{a}$ and $W^{ \prime}$ 
are the strength fields, and they are given by 

\begin{eqnarray}
W^{a}_{\alpha c}&=&- \frac{1}{8g} \bar{D} \bar{D} e^{-2g \hat{V}_{c}} 
D_{\alpha} e^{-2g \hat{V}_{c}} \nonumber \\
W^{a}_{\alpha}&=&- \frac{1}{8g} \bar{D} \bar{D} e^{-2g \hat{V}} 
D_{\alpha} e^{-2g \hat{V}} \nonumber \\
W^{\prime}_{\alpha}&=&- 
\frac{1}{4} \bar{D} \bar{D} D_{\alpha} \hat{V}^{\prime} \,\ .
\label{l6}
\end{eqnarray}

Finally 
\begin{eqnarray}
  {\cal L}_{Scalar} 
     &=&  \int d^{4}\theta\;\left[\,\hat{ \bar{ \eta}}e^{2g\hat{V}}   
                         \hat{ \eta} +			  
\hat{ \bar{ \rho}}e^{2g\hat{V}+g'\hat{V}'}                            
\hat{ \rho} +
\hat{ \bar{ \chi}}e^{2g\hat{V}-g'\hat{V}'}                            
\hat{ \chi} + 
\hat{ \bar{S}}e^{2g\hat{V}}                            
\hat{S}\right.
\nonumber \\ &+&
\left. \hat{ \bar{ \eta}}^\prime e^{2g\hat{\bar V}}   
                         \hat{ \eta}^\prime +			  
\hat{\bar{ \rho}}^\prime e^{2g\hat{\bar V}-g'\hat{\bar V}'}                     
\hat{ \rho}^\prime +
\hat{ \bar{ \chi}}^\prime e^{2g\hat{\bar V}+g'\hat{\bar V}'}                    
\hat{ \chi}^\prime + 
\hat{ \bar{S}}^\prime e^{2g\hat{\bar V}}                            
\hat{S}^\prime\right]\nonumber \\ &+&
\int d^2\theta\, W+\int d^2\bar\theta\, \bar{W},
\label{l7}
\end{eqnarray}
where $W$ is the superpotential, which we discuss in the next subsection. 

\subsection{Superpotential.}
\label{subsec:spotential}

The superpotential of our model is given by
\begin{equation}
W=\frac{W_{2}}{2}+ \frac{W_{3}}{3}, 
\label{sp1}
\end{equation}
with $W_{2}$ having only two chiral superfields and the terms permitted by  
our symmetry are
\begin{equation}
W_{2}=\mu_{0}\hat{L} \hat{ \eta}^\prime+\mu_{ \eta} \hat{ \eta} 
\hat{ \eta}^\prime+
\mu_{ \rho} \hat{ \rho} \hat{ \rho}^\prime+
\mu_{ \chi} \hat{ \chi} \hat{ \chi}^\prime+
\mu_{S} \hat{S} \hat{S}^\prime,
\label{sp2}
\end{equation}
and in the case of three chiral superfields the terms are
\begin{eqnarray}
W_{3}&=& \lambda_{1} \epsilon \hat{L} \hat{L} \hat{L}+
\lambda_{2} \epsilon \hat{L} \hat{L} \hat{ \eta}+
\lambda_{3} \hat{L} \hat{L}
\hat{S}+\lambda_4\epsilon\hat{L}\hat{\chi}\hat{\rho}+
f_{1} \epsilon \hat{ \rho} \hat{ \chi} \hat{ \eta}+
f_{2} \hat{ \eta} \hat{ \eta} \hat{S}+
f_{3} \hat{ \chi} \hat{ \rho} \hat{S}+f'_1 \epsilon \hat{ \rho}^\prime 
\hat{ \chi}^\prime \hat{ \eta}^\prime \nonumber \\
&+& f'_2 \hat{ \eta}^\prime \hat{ \eta}^\prime \hat{S}^\prime
+ f'_3 \hat{ \chi}^\prime \hat{ \rho}^\prime \hat{S}^\prime+\sum_i[ \kappa_{1i}
\hat{Q}_{1} \hat{\eta}^{\prime} \hat{u}^{c}_{i}+ 
\kappa_{2i} \hat{Q}_{1} \hat{\rho}^{\prime} \hat{d}^{c}_{i}]+
\kappa_{3} \hat{Q}_{1} \hat{\chi}^\prime \hat{J}^{c}+
\sum_{\alpha i} [\kappa_{4\alpha i} \hat{Q}_{\alpha} \hat{\eta} 
\hat{d}^{c}_{i}\nonumber \\  &+&
\kappa_{5\alpha i} \hat{Q}_{\alpha} \hat{\rho} \hat{u}^{c}_{i}]+
\sum_{\alpha \beta} 
\kappa_{6\alpha\beta}\hat{Q}_{\alpha } \hat{\chi} \hat{j}^{c}_{\beta} 
+ \sum_{\alpha ij}\kappa_{7\alpha ij} \hat{Q}_{\alpha} \hat{L}_{i} 
\hat{d}^{c}_{j}+
\sum_{i,j,k}\xi_{1ijk} \hat{d}^{c}_{i} \hat{d}^{c}_{j} \hat{u}^{c}_{k}
\nonumber \\ &+&
\sum_{ij\beta}
\xi_{2ij\beta} \hat{u}^{c}_{i} \hat{u}^{c}_{j} \hat{j}^{c}_{\beta}+
\sum_{i\beta}
\xi_{3 i\beta} \hat{d}^{c}_{i} \hat{J}^{c} \hat{j}^{c}_{\beta},
\label{sp3}
\end{eqnarray}
with $i,j,k=1,2,3$, $\alpha=2,3$ and $\beta=1,2$.

All the eight neutral scalar components 
$\eta^0,\rho^0,\chi^0,\sigma^0_2,\eta^{\prime0},\rho^{\prime0},
\chi^{\prime0}$
gain non-zero vacuum expectation values. 
This arises from the mass matrices for quarks. In fact, defining $\langle
\eta^0\rangle=v_\eta/\sqrt2$, $\langle 
\eta^{0\prime}\rangle =v^{\prime}_\eta/\sqrt2$, etc,
from the superpotential in Eq.~(\ref{sp3}) the following mass
matrices arise
\begin{equation}
\Gamma^u=\frac{1}{\sqrt2}\,\left(\begin{array}{ccc}
\kappa_{11}v^{\prime}_\eta & \kappa_{12}v^{\prime}_\eta &
\kappa_{13}v^{\prime}_\eta\\ 
\kappa_{521}v_\rho & \kappa_{522}v_\rho & \kappa_{523}v_\rho \\
\kappa_{531}v_\rho & \kappa_{532}v_\rho & \kappa_{533} v_\rho\\
\end{array}\right),
\label{qumm}
\end{equation}
for the $u$-quarks, and 
\begin{equation}
\Gamma^d=\frac{1}{\sqrt2}\,\left(\begin{array}{ccc}
\kappa_{21}v^{\prime}_\rho & \kappa_{22} v^{\prime}_\rho &
\kappa_{23}v^{\prime}_\rho \\ 
\kappa_{421} v_\eta  & \kappa_{422} v_\eta & \kappa_{423} v_\eta \\
 \kappa_{431} v_\eta& \kappa_{432} v_\eta & \kappa_{4333} v_\eta\\
\end{array}\right),
\label{qdmm}
\end{equation}
for the $d$-quarks, and for the exotic quarks, $J$ and $j_{1,2}$,we have
$M_J=\kappa_3v^\prime_\chi$ and 
\begin{equation}
\Gamma^j=\frac{v_\chi}{\sqrt2}\left(\begin{array}{cc}
\kappa_{621} & \kappa_{622} \\
\kappa_{631} & \kappa_{632}
\end{array}\right),
\label{qjmm}
\end{equation}
respectively.

From Eqs.(\ref{qumm}) and (\ref{qdmm}) we see that all the VEVs have
to be different from zero in order to give mass to all quarks. Notice also that
the $u$-like and $d$-like mass matrices have no common VEVs. On the other
hand, the charged lepton mass matrix is already given by
$M^l_{ij}=v_{\sigma_2}\lambda_{3ij}/\sqrt2$, where $v_{\sigma_2}$ is the VEV of
the $\langle\sigma^0_2\rangle$ component of the anti-sextet $S$ in
Eq.(\ref{sextet}). However, $v^{\prime}_{\sigma_{1,2}}$ can both be zero since
the sextet $S^\prime$ does not couple to leptons at all. 

The terms with $\mu_0,\xi_i$ and $f_2$ in the superpotential $W_3$ given in
Eq.~(\ref{sp3}) violate the conservation of the ${\cal F}=B+L$
quantum number. For instance, if we allow the $\xi_1$ term it implies 
in proton decay~\cite{pal2}. However, if we assume the global $U(1)_{\cal F}$ 
symmetry, it allows us to introduce the $R$-conserving symmetry~\cite{rs},
defined as $R=(-1)^{3{\cal F}+2S}$.
The ${\cal F}$ number attribution is
\begin{equation}
\begin{array}{c}
{\cal F}(U^{--})={\cal F}(V^{-}) = - {\cal F}(J_1)= {\cal F}(J_{2,3})=
{\cal F}(\rho^{--})  \\  
= {\cal F}(\chi^{--}) ={\cal F}(\chi^{-}) = 
{\cal F}(\eta^-_2)={\cal F}(\sigma_1^0)=2,\end{array}
\label{efe}
\end{equation}
with ${\cal F}=0$ for the other Higgs scalar, while for leptons and the known 
quarks ${\cal F}$ coincides with the total lepton and baryon numbers, 
respectively. As in the MSSM this definition implies that all known
standard model's particles have even $R$-parity while their supersymmetric
partners have odd $R$-parity. The terms  
which are proportional to the following constants: $\mu_0$ in Eq.~(\ref{sp2});
$\lambda_1,\lambda_4,f_2,f^\prime_2,\kappa_7,\xi_{1,2,3}$ in Eq.~(\ref{sp3})
violate the $R$-parity defined above. 
The terms $\xi_2,\lambda_4$ were not considered in Ref.~\cite{pal2}.
However, the term with $\xi_2$ involves an exotic quark (heavier than the
proton) so the analysis in that reference is still valid. 

As usual, the supersymmetry breaking is accomplish by including the most
general renormalizable soft-supersymmetry breaking terms but now, they must
be also consistent with the 3-3-1 gauge symmetry. We will also include terms
which explicitly violate the $R$-like symmetry.
These soft terms are given by
\begin{eqnarray}
{\cal L}_{soft}&=&- \frac{1}{2} \left[m_{\lambda_c}\sum_a\lambda^a_c\lambda^a_c
+ m_{ \lambda} \sum_a \left( \lambda^{a} \lambda_{a} \right) 
+m^{ \prime} \lambda \lambda+H.c. \right]-m_{L}^{2} \tilde{L}^{ \dagger} 
\tilde{L}-m_{Q_1}^{2} \tilde{Q}^{ \dagger}_{1} \tilde{Q}_{1}\nonumber \\ &+&
\sum_i[ \tilde{u}^{\dagger}_{i} m_{u_{i}}^2
\tilde{u}_{i}-
\tilde{d}^{\dagger}_{i}m_{d_{i}}^2 \tilde{d}_{i}]-
m_{J}^{2} \tilde{J}^{\dagger} \tilde{J}-
\sum_\beta \tilde{j}^{\dagger}_{\beta}
m_{j_{ \beta}}^{2} \tilde{j}_{ \beta}-
\sum_\alpha m_{Q_{\alpha}}^{2} \tilde{Q}^{ \dagger}_{\alpha} \tilde{Q}_{\alpha}
\nonumber \\ &- &m_{ \eta}^{2} \eta^{ \dagger} \eta 
-m_{ \chi}^{2} \chi^{ \dagger} \chi -
m_{ \rho}^{2} \rho^{ \dagger} \rho-m_s^{2}{\rm Tr}\, (S^{ \dagger}
S)-m^2_{\eta'}\eta^{\prime\dagger}\eta^\prime-
m^2_{\rho'}\rho^{\prime\dagger}\rho^\prime- 
m^2_{\chi'}\chi^{\prime\dagger}\chi^\prime-m^2_{s'}\,{\rm Tr}
(S^{\prime\dagger}S^\prime)
\nonumber \\ &+& \left[ k_{1} \epsilon \rho \chi \eta+
k_{2} \eta \eta S^{ \dagger}+ k_{3} \chi \rho S^{ \dagger}+
k^\prime_{1} \epsilon \rho^\prime \chi^\prime \eta^\prime+
k^\prime_{2} \eta^\prime \eta^\prime S^{\prime \dagger}+ 
k^\prime_{3} \chi^\prime \rho^\prime S^{\prime \dagger}
 + \right. 
\nonumber\\ &+&  
\left.-M^2\tilde{L} \eta^{ \dagger}+
\varepsilon_0\sum_{ijk}\epsilon_{ijk}\tilde{L}_i\tilde{L}_j\tilde{L}_k+
\varepsilon_1\sum_{ijk}\epsilon_{ijk}\tilde{L}_i\tilde{L}_j\eta_k
+\varepsilon_2\sum_{ij}\tilde{L}_i\tilde{L}_j S_{ij}+\varepsilon_3
\sum_{ijk}\epsilon_{ijk}\tilde{L}_i\chi_j\rho_k \right.
\nonumber \\
&+& \left. \sum_i\tilde{Q}_{1}\left(\zeta_{1i}  \eta^\prime \tilde{u}^{c}_{i}+
\zeta_{2i}  \rho^\prime \tilde{d}^{c}_{i}+\zeta_{3J}\chi^\prime
\tilde{J}^{c}\right)+ 
\sum_{\alpha } 
\tilde{Q}_{\alpha}(\sum_i(\omega_{1\alpha i} \eta \tilde{d}^{c}_{i}+
\omega_{2\alpha i} \rho \tilde{u}^{c}_{i}+
\sum_j\omega_{3 \alpha ij}  \tilde{L}_{i} \tilde{d}^{c}_{j})\right.
\nonumber \\
&+& \left.\sum_\beta\omega_{4\alpha \beta}  \chi \tilde{j}^{c}_{\beta})  +
\sum_{i}\left(\sum_{jk}\varsigma_{1ijk} \tilde{d}^{c}_{i} \tilde{d}^{c}_{j} 
\tilde{u}^{c}_{k}+ 
\sum_\alpha \left[\varsigma_{2i\alpha} \tilde{d}^{c}_{i} \tilde{J}^{c} 
\tilde{j}^{c}_{\alpha}+
\sum_j\varsigma_{3ij\alpha} \tilde{u}^{c}_{i} \tilde{u}^{c}_{j} 
\tilde{j}^{c}_{\alpha}\right]\right) +H.c.\right].
\label{soft}
\end{eqnarray}
The terms proportional to
$k_2,k^\prime_2,M^2,\varepsilon_0,\varepsilon_3,\omega_3,\varsigma_{1,2,3}$
violates the $R$-parity symmetry. 

The $SU(3)$ invariance tell us that in
\begin{equation}
m_{L} = 
      \left( \begin{array}{ccc} 
m_{\tilde{\nu}}& 0& 0 \\ 
0& m_{\tilde{l}}& 0 \\
0& 0&  m_{\tilde{l}^{c}}
\end{array} \right) \,\ ,
\label{mm}
\end{equation}
we need $m_{\tilde{l}^{c}}=m_{\tilde{l}}\not=m_{\tilde{\nu}}$ and the same
for the other mass parameters. More details of the lagrangian will be
given elsewhere~\cite{mcr}. 
\section{The scalar potential}
\label{sec:sp}

The scalar potential is written as
\begin{mathletters}
\label{potential}
\begin{equation}
V_{331}=V_D+V_F+V_{\mbox{soft}}
\label{ep1}
\end{equation}
where

\begin{eqnarray}
V_D&=&-{\cal L}_D=\frac{1}{2}\left(D^aD^a+DD\right)\nonumber \\ &=&
\frac{g^{\prime2}}{2}(\rho^\dagger\rho-\rho^{\prime\dagger}\rho^\prime
-\chi^\dagger\chi+\chi^{\prime\dagger}\chi^\prime)^2+
\frac{g^2}{8}\sum_{i,j}\left(\eta^\dagger_i\lambda^a_{ij}\eta_j
+\rho^\dagger_i\lambda^a_{ij}\rho_j
+\chi^\dagger_i\lambda^a_{ij}\chi_j+S^\dagger_{ij}\lambda^a_{jk}
S_{kl}\right.
\nonumber \\ &-&
\left.\eta^{\prime\dagger}_i\lambda^{*a}_{ij}\eta^\prime_j 
-\rho^{\prime\dagger}_i\lambda^{*a}_{ij}\rho^\prime_j
-\chi^{\prime\dagger}_i\lambda^{*a}_{ij}\chi^\prime_j-
S^{\prime\dagger}_{ij}\lambda^{*a}_{jk} S^\prime_{kl} \right)^2,
\label{esd}
\end{eqnarray}

\begin{eqnarray}
V_F&=&-{\cal L}_F=\sum_mF^*_m F_m\nonumber \\ &=&
\sum_{i,j,k}\left[\left\vert \frac{\mu_\eta}{2} \eta^{\prime}_i+\frac{f_1}{3}
\epsilon_{ijk}\rho_j\chi_k
+\frac{2f_2}{3}\eta_iS_{ij}\right\vert^2+
\left\vert \frac{\mu_\rho}{2}\rho^\prime_i+\frac{f_1}{3}
\epsilon_{ijk}\chi_j\eta_k+\frac{f_3}{3}\chi_i S_{ij}\right\vert^2
\right.\nonumber \\ &+&\left.
\left\vert
\frac{\mu_\chi}{2}\chi^\prime_i+\frac{f_1}{3}\epsilon_{ijk}\rho_j\eta_k 
+\frac{f_3}{3}\rho_i S_{ij}\right\vert^2
+\left\vert \frac{\mu_s}{2}S^\prime_{ij}+\frac{f_2}{3}\eta_i\eta_j+\frac{f_3}{3}
\chi_i\rho_j\right\vert^2\right.\nonumber \\ &+&\left.\left\vert 
\frac{\mu_\eta}{2} \eta_i+\frac{f^\prime_1}{3}
\epsilon_{ijk}\rho^\prime_j\chi^\prime_k
+\frac{2f^\prime_2}{3}\eta^\prime_iS^\prime_{ij}\right\vert^2+
\left\vert \frac{\mu_\rho}{2}\rho_i+\frac{f^\prime_1}{3}
\epsilon_{ijk}\chi^\prime_j\eta^\prime_k+\frac{f^\prime_3}{3}\chi^\prime_i
S^\prime_{ij} 
\right\vert^2\right.\nonumber \\ &+&\left.
\left\vert
\frac{\mu_\chi}{2}\chi_i+\frac{f^\prime_1}{3}\epsilon_{ijk}\rho^\prime_j
\eta^\prime_k
+\frac{f^\prime_3}{3}\rho^\prime_i S^\prime_{ij}\right\vert^2
+\left\vert
\frac{\mu_s}{2}S_{ij}+\frac{f^\prime_2}{3}\eta^\prime_i\eta^\prime_j+ 
\frac{f^\prime_3}{3}
\chi^\prime_i\rho^\prime_j\right\vert^2
\right],
\label{esf}
\end{eqnarray}

\begin{eqnarray}
V_{soft}&=&-{\cal L}^{\mbox{scalar}}_{soft}\nonumber \\ &=&
m^2_\eta\eta^\dagger\eta+m^2_\rho\rho^\dagger\rho+m^2_\chi\chi^\dagger\chi
+m^2_s\,{\rm Tr} (S^\dagger S)+
m^2_{\eta'}\eta^{\prime\dagger}\eta^\prime+
m^2_{\rho'}\rho^{\prime\dagger}\rho^\prime
+m^2_{\chi'}\chi^{\prime\dagger}\chi^\prime
+m^2_{s'}{\rm Tr}\,( S^{\prime\dagger} S^\prime)\nonumber \\ &+&
[k_1\epsilon\,\rho\chi\eta+k_2\eta S^\dagger \eta 
+k_3\chi^TS^\dagger\rho +
k^\prime_1\epsilon\,\rho^\prime\chi^\prime\eta^\prime+k^\prime_2
\eta^\prime S^{\prime\dagger} \eta^\prime
+k^\prime_3\chi^{\prime T}S^{\prime\dagger}\rho^\prime+H.c.].
\label{ess}
\end{eqnarray}
\end{mathletters}
Note that $k_{1,2,3},k^\prime_{1,2,3}$ has dimension of mass and that 
the terms which are proportional to $k_2,k^\prime_2$ and $f_2,f^\prime_2$
violate the $R$-parity.

It is instructive to rewrite Eqs.(\ref{esd}) as follows
\begin{eqnarray}
V_D&=&\frac{{g'}^2}{2}\left(\rho^\dagger\rho-\chi^\dagger\chi
-\rho^{\prime\dagger}\rho^\prime+\chi^{\prime\dagger}\chi^\prime\right)^2
+\frac{g^2}{8}\left[\frac{4}{3}\sum_i (X^\dagger_i X_i)^2+
2\sum_{i,j}(X^\dagger_i X_j) (X^\dagger_j X_i)\right.\nonumber \\
&+&\left.2\sum_i (X^\dagger_i 
S)(S^\dagger X_i)-  
\frac{4}{3}\sum_{i,j}(X^\dagger_i X_i)(X^\dagger_j X_j)
-\frac{2}{3}{\rm Tr}(S^\dagger S)\sum_i(X^\dagger_i X_i)\right.\nonumber \\
&+&\left.
2{\rm Tr}[(S^\dagger S)^2]-\frac{2}{3}({\rm Tr} S^\dagger S)^2+{\rm pt}
\right],
\label{esd2}
\end{eqnarray} 
where ``pt'' in the expression above denotes the replacements 
$X^\prime_i\leftrightarrow X_i$ but not in $g^\prime$ which is
always the coupling constant of the $U(1)_N$ factor. In the same way we rewrite
Eq.~(\ref{esf}) as  
\begin{eqnarray}
V_F&=&\sum_i\frac{\mu^2_{X_i}}{4}( X^\dagger_i X_i+{\rm pt}) +\frac{\vert
\mu^2_S\vert}{4} 
({\rm Tr}(S^\dagger S)+{\rm pt})+\frac{1}{6} 
\sum_{i\not=j\not=k}\{\frac{\mu^*_{X_i}}{6}[(f_1\epsilon_{ijk}
X^{\prime\dagger}_iX_j 
X_k+f^\prime_1\epsilon_{ijk}X^{\dagger}_i X^\prime_j X^\prime_k ] 
\nonumber\\ &+& 
\frac{f_2}{3}[\mu^*_\eta \eta^{\prime\dagger} S
\eta+\frac{\mu^*_S}{2}\eta^\dagger S^\prime\eta]+{\rm pt}+
\frac{f_3}{6}[\mu^*_\rho\rho^{\prime\dagger}S\chi+\mu^*_\chi\chi^{\prime\dagger}
S\rho+\mu^*_S\chi^\dagger S^\prime \rho]+{\rm pt}+H.c.\} 
\nonumber \\ &+&\vert f_1\vert^2[\sum_{i\not=j}
(X^\dagger_iX_i)(X^\dagger_j X_j)-(X^\dagger_j X_i)(X^\dagger_i X_j)]+{\rm pt}
+\frac{4\vert f_2 \vert^2}{9}[(\eta S)^\dagger(\eta S)+(\eta^\dagger\eta)^2]
+{\rm pt} 
\nonumber \\ &+&
\frac{\vert f_3\vert^2}{9}[(\chi S)^\dagger(\chi S)+(\rho S)^\dagger(\rho S)+
(\chi^\dagger\chi)^\dagger(\rho^\dagger\rho)]+{\rm pt}+\frac{2f_1^*f_2}{9}
\epsilon(\rho\chi)^\dagger\eta S+{\rm
pt}\nonumber \\ &+&[\frac{f_1^*f_3}{9}\epsilon(\chi\eta)^\dagger\chi S+{\rm pt}
+\frac{f^*_2f_3}{9}(\eta^\dagger\chi)(\eta^\dagger\rho)+{\rm pt}+H.c.], 
\label{esf2}
\end{eqnarray}
where ``pt'' in the expression above denotes as before the replacements 
$X^\prime_i\leftrightarrow X_i$, and $f^\prime_{1,2,3}\leftrightarrow
f_{1,2,3}$, 
and $k^\prime_{1,2,3}\leftrightarrow k_{1,2,3}$. We have omitted $SU(3)$ indices
since we have denoted the unprimed triplets wherever it is possible as
$X_i=\eta,\rho,\chi$ and $X^\prime_i=\eta^\prime,\rho^\prime,\chi^\prime$ but in
each term only unprimed (primed) field appears. 

We can now work out the mass spectra of the scalar and pseudoscalar fields by 
making a shift of the form $X \to \frac{1}{\sqrt{2}}(v_X +H_X+iF_X)$ (similarly
for the case of the primed fields)
for  all the neutral scalar fields of the multiplets $X_i$. Note that $H_X$ and
$F_X$ are not mass eigenstates yet. We will denote $H_i,\,i=1,\cdots8$ and 
$A_i,\,i=1,\cdots,6$ the respective massive fields; $G_{1,2}$ will denote the two
neutral Goldstone bosons.  
The mass matrices appear in the Appendix~\ref{sec:a1}, for the real scalars, and
in the Appendix~\ref{sec:a2} for the pseudoscalar case. The constraint
equations are given in the Appendix~\ref{sec:a3}. 
We will use below the following set of parameters in the scalar potential:
\begin{equation}
f_1=f_3=1,\quad f^\prime_1=f^\prime_3=10^{-6},\quad{\rm (dimensionless)}
\label{fs}
\end{equation}
and
\begin{equation}
-k_1=k^\prime_1=10,\;k_3=k^\prime_3=-100,\;  
-\mu_\eta=\mu_\rho=- \mu_s= \mu_\chi=1000, \quad \mbox{(in GeV)},
\label{ks}
\end{equation}
we also use the constraint 
$V^2_\eta+V^2_\rho+2V^2_2=(246\;{\rm GeV})^2$ coming from $M_W$, where,
we have defined $V^2_\eta=v^2_\eta+v^{\prime2}_\eta$ and $V^2_\rho=
v^2_\rho+v^{\prime2}_\rho$ and $V^2_2=v^2_{\sigma_2}+v^2_{\sigma^\prime_2}$.
Assuming that $v_\eta=20$, $v_\chi=1000$, $v_{\sigma_2}=10$, 
$v^\prime_\eta=v^\prime_\rho=v^\prime_{\sigma_2}=v^\prime_\chi=1$ in GeV, the
value of $v_\rho$ is fixed by the constraint above. 

With this set of values for the parameters
the real mass eigenstates $H_i$ are obtained by the diagonalization
of the mass matrix given in the Appendix~\ref{sec:a1}. 
Besides the constraint equations (Appendix \ref{sec:a3}) and imposing the
positivity of the eigenvalues (mass square), and the values for the parameters
given above we obtain the following values for
the masses of the scalar sector (in GeV)
$M_{H_i}=121.01, 277.14, 515.26, 963.68, 1218.8, 1243.24,3797.86,4516.43$, 
$i=1,\cdots,8$ and $M_{H_j}>M_{H_i}$ with $j>i$. 
In the pseudoscalar sector we have verified analytically that the mass matrix in
the Appendix~\ref{sec:a2}, has two
Goldstone bosons as it should. The other six physical pseudoscalars have the
following masses, with the same parameters as before, in GeV, 
$M_{A_i}=276.4,515.3,963.65,1243.24,3797.85,4516.43$. 

The behavior of the lightest scalar ($H_1$)and pseudoscalar ($A_1$) as a 
function of $v_\chi$ is shown in Fig.~1 for a given  
choice of the parameters, we see that, at the tree level, there is an upper
limit for the mass of the lightest scalar: $M_{H_1}<124.5$ GeV and that
for these values of  the parameters $M_{A_1}>M_Z$. Other values of the
parameters give higher or lower values for the upper limit of $M_{H_1}$.
Of course, radiative corrections have to be taken into account, however, this 
has to be done in the context of the supersymmetric 3-3-1 model which is not in
the scope of the present work. Hence the mass square of the lightest real scalar
boson has an upper bound (see Fig.~1) 
\begin{equation}
M^2_{H_1}\leq (124.5+\epsilon)^2\;{\rm GeV}^2
\label{ub1}
\end{equation}
where 124.5 GeV is the tree value ($\epsilon=0$).
We recall that in the MSSM if $M_{A_1}>M_Z$ the upper limit on the
mass of the lightest neutral scalar is $M_Z$ at the tree level but 
radiative corrections rise it to 130 GeV~\cite{haber2}. 

\section{Conclusions}
\label{sec:con}

We have built the complete supersymmetric version of the 3-3-1 model of
Refs.~\cite{pp,pf}. Another possibility in this 3-3-1 model 
which avoids the introduction of the scalar sextet, $S$, was considered some
years ago by Duong and Ma, Ref.~\cite{ema1}, who built the
supersymmetric version of that model. The sextet was substituted by a single
charged lepton singlet $E_L\sim({\bf1},1)$ and $E^c_L\sim({\bf1},-1)$.
Here we would like to point out the differences between our version of 
the supersymmetric 3-3-1 model and that of Ref.~\cite{ema1}. 
{\bf a)} Duong and Ma assumed that the breaking of $SU(3)_L\otimes U(1)_N\to
SU(2)_L\otimes U(1)_Y$ occurred before the breaking of supersymmetry and the
resulting model is a supersymmetric $SU(2)_L\otimes U(1)_Y$ model. 
Even, in this case the scalar potential involving doublets of the residual gauge
symmetry do not coincide with the potential of the MSSM. In the present work, we
have considered that the supersymmetry is broken at the same time that the 3-3-1
gauge symmetry. Hence, we have to consider the complete 3-3-1 scalar potential.
It means that in the Duong and Ma supersymmetric model there are no doubly
charged charginos and exotic charged squarks.
{\bf b)} In Ref.~\cite{ema1} it was assumed that some of the VEVs have zero
value, unlikely we have considered all (but $\sigma_2^{\prime0}$ and 
$\sigma_1^{0}$) of them different from zero. 
Hence we are able to obtain realistic quark and charged lepton masses, as can be
seen from Eqs.(\ref{qumm}), (\ref{qdmm}) and (\ref{qjmm}). In Ref.~\cite{ema1}
some of these masses have to be generated by radiative corrections~\cite{ema2}.
From a) and b) we see that the 3-3-1 supersymmetric model considered in this
work has different phenomenological features from the supersymmetric 3-3-1 model
of Duong and Ma.
  
From the phenomenological point of view there are several possibilities.
Since it is possible to define the $R$-parity symmetry, the phenomenology 
of this model with $R$-parity conserved has similar features to that of the 
$R$-conserving MSSM: the supersymmetric particles are pair-produced and the
lightest neutralino is the lightest supersymmetric particle (LSP). 
The mass spectra of all particles in this model will be considered
elsewhere~\cite{mcr}. However, there are differences between
this model and the MSSM with or without $R$-parity breaking:
due to the fact that there are doubly charged scalar and vector fields. 
Hence, we have doubly charged charginos which are mixtures of the
superpartners of the $U$-vector boson with the doubly charged higgsinos. 
This implies new interactions that are not present in the 
MSSM, for instance: $\tilde{\chi}^{--}\tilde{\chi}^0U^{++}$,
$\tilde{\chi}^-\tilde{\chi}^-U^{++}$, 
$\tilde{l}^-l^-\tilde{\chi}^{++}$ where $\tilde{\chi}^{++}$ denotes
any doubly charged chargino. 
Moreover, in the chargino production, besides the usual mechanism, we have 
additional contributions coming from the $U$-bilepton in the s-channel. 
Due to this fact we expect that there will be an enhancement in the cross 
section of production of these particles in $e^-e^-$ collisors, such as the
NLC~\cite{mcr}. 
We will also have the singly charged charginos and neutralinos, as in 
the MSSM, where there are processes like $\tilde{l}^-l^+\tilde{\chi}^0$, 
$\tilde{\nu_l}l^-\tilde{\chi}^{+}$, with $\tilde{l}$ denoting any 
slepton; $\tilde{\chi}^-$ 
denotes singly charged chargino and $\tilde{\nu}_L$ denotes any 
sneutrino. The only difference is that in the MSSM there are five neutralinos
and in the 3-3-1s model there are eight neutralinos. 

Finally, we would like to call attention that, whatever the energy scale $\mu$
at which $\sin^2\theta_W(\mu)=1/4$ is in the non-supersymmetric 3-3-1 model,
when supersymmetry is added it will result a rather different value for $\mu$.
In conclusion, we can say that the present model has a rich phenomenology that
deserves to be studied more in detail.
     
\acknowledgments 
This work was supported by Funda\c{c}\~ao de Amparo \`a Pesquisa
do Estado de S\~ao Paulo (FAPESP), Conselho Nacional de 
Ci\^encia e Tecnologia (CNPq) and by Programa de Apoio a
N\'ucleos de Excel\^encia (PRONEX).
One of us (MCR) would like to thank the Laboratoire de Physique 
Math\'ematique et Th\'eorique, Universit\'e  Montpellier II, 
for its kind hospitality 
and also M. Capdequi-Peyran\`ere and G. Moultaka for useful discussions.

\appendix

\section{Mass matrix of the scalar neutral fields}
\label{sec:a1}
Here we write down the complete symmetric mass matrix in the scalar CP-even 
sector, the constraint equation given in the Appendix~\ref{sec:a3} have been
already taken into account.

\begin{eqnarray*}
&&{\bf M_{11}}= \frac{g^2v^2_\eta}{3}+ \frac{1}{18\sqrt{2}v_\eta}
(f_1f_3v^2_\rho v_{\sigma_2}-18k_1v_\rho v_\chi+3f_1v^\prime_\rho \mu_\rho
v_\chi+f_1f_3v_{\sigma_2}v^2_\chi-3f^\prime_1\mu_\eta v^\prime_\rho 
v^\prime_\chi+3f_1\mu_\chi v_\rho v^\prime_\chi),
\nonumber \\&& {\bf M_{12}}=-\frac{g^2v_\eta
v_\rho}{6}+\frac{1}{9\sqrt2}(\sqrt{2}f^2_1 v_\eta v_\rho-f_1f_3v_\rho
v_{\sigma_2}+9k_1 v_\chi-\frac{3}{2}\mu_\chi v^\prime_\chi),
\nonumber \\&& {\bf M_{13}}=-\frac{g^2v_\eta
v_\rho}{6}+\frac{1}{9\sqrt2}(9k_1v_\rho- 
\frac{3}{2}f_1\mu_\rho v^\prime_\rho+\sqrt{2}f^2_1v_\eta
v_\chi-f_1f_3v_{\sigma_2}v_\chi)\nonumber \\&&
{\bf M_{14}}=\frac{g^2v_\eta v_{\sigma_2}}{6}-
\frac{f_1f_3}{18\sqrt2}(v^2_\rho+v^2_\chi),\;
{\bf M_{15}}=-\frac{g^2v_\eta v^\prime_\eta}{3},\; 
{\bf M_{16}}=\frac{g^2 v_\eta v^\prime_\rho}{6}-
\frac{1}{6\sqrt2}(\mu_\rho v_\chi-\mu_\eta v^\prime_\chi),\nonumber \\&&
{\bf M_{17}}=\frac{g^2 v_\eta v^\prime_\chi}{6}+
\frac{1}{6\sqrt2}(f^\prime_1 \mu_\eta v^\prime_\rho-f_1\mu_\chi v_\rho)
,\;{\bf M_{18}}=\frac{g^2 v_\eta v_{\sigma_2}}{6} \nonumber \\&&
{\bf M_{22}}=(\frac{g^2}{3}+g^{\prime 2})v^2_\rho-
\frac{v_\chi}{12\sqrt2 v_\rho}(12k_1v_\eta+ 6\sqrt2 k_3v_{\sigma_2}+
2f_1\mu_\eta v^\prime_\eta+\sqrt2 f_3\mu_s v^\prime_{\sigma_2})
\nonumber \\ &+&\frac{v^\prime_\chi}{12\sqrt2 v_\rho}(2f^\prime_1\mu_\rho
v^\prime_\eta-\sqrt2 
f^\prime_3\mu_\rho v^\prime_{\sigma_2}+2f_1\mu_\chi v_\eta-\sqrt2 \mu_\chi
v_{\sigma_2}v^\prime_\chi),
\nonumber \\&&
{\bf M_{23}}=-(\frac{g^2}{6}+g^{\prime2})v_\rho v_\chi+\frac{1}{12\sqrt2}(12k_1v_\eta+
6\sqrt2 k_3v_{\sigma_2}+2f^\prime \mu_\eta v^\prime_\eta+\sqrt2 f_3\mu_s
v^\prime_{\sigma_2})\nonumber \\ &+&\frac{v_\rho v_\chi}{9}(f^2_1+f^2_3),\;
{\bf M_{24}}=\frac{g^2}{12}v_\rho v_\chi+\frac{1}{18\sqrt2}(-2f_1f_3 v_\eta v_\rho+
\sqrt2 f^2_3 v_{\sigma_2}v_\rho+9\sqrt2 k_3 v_\chi+\frac{3}{\sqrt2}f_3\mu_\chi
v^\prime_\chi),
\nonumber \\&& {\bf M_{25}}=\frac{g^2}{6}v^\prime_\eta
v_\rho+\frac{1}{6\sqrt2}(f_1\mu_\eta v_\chi-f^\prime_1\mu_\rho v^\prime_\chi),\;
{\bf M_{26}}=-(\frac{g^2}{3}+g^{\prime2})\frac{v_\rho v^\prime_\rho}{3},
\nonumber \\&&
{\bf M_{27}}=(\frac{g^2}{6}+g^{\prime2})v_\rho
v^\prime_\chi-\frac{\mu_\rho}{12\sqrt2}(2f^\prime_1 v^\prime_\eta-f^\prime_3
v^\prime_{\sigma_2})-\frac{\mu_\chi}{12\sqrt2}(2f_1v_\eta-f_3v_{\sigma_2}),
\nonumber \\&&
{\bf M_{28}}=-\frac{g^2v_\rho v^\prime_{\sigma_2}}{12}+\frac{1}{12}(f_3\mu_s
v_\chi+f^\prime_3\mu_\rho v^\prime_\chi),\nonumber \\ &&
{\bf M_{33}}=(\frac{g^2}{3}+g^{\prime2})v^2_\chi-\frac{1}{12\sqrt2 v_\chi}(12k_1v_\eta
v_\rho+\frac{12}{\sqrt2}k_3v_\rho v_{\sigma_2}+2f_1\mu_\eta v^\prime_\eta
v_\rho-2f_1\mu_\rho v_\eta v^\prime_\rho+\sqrt2f_3\mu_\rho v^\prime_\rho
v_{\sigma_2}\nonumber \\ &+&\sqrt2f_3\mu_s v_\rho
v^\prime_{\sigma_2}-2f^\prime_1\mu_\chi v^\prime_\eta
v^\prime_\rho+\sqrt2f_3\mu_\chi v^\prime v^\prime_{\sigma_2}),
\;
{\bf M_{34}}=\frac{g^2}{12}v_{\sigma_2}v_\chi+\frac{1}{18\sqrt2}
(\frac{9}{\sqrt2}k_3v_\rho+\frac{3}{\sqrt2}f_3\mu_\rho
v^\prime_\rho\nonumber\\&-&2f_1f_3v_\eta v_\chi+\sqrt2f^2_3
v_{\sigma_2}v_\chi),\; {\bf M_{35}}=\frac{g^2}{6}v^\prime_\eta
v_\chi+\frac{1}{6\sqrt2}(f_1\mu_\rho v_\rho-f^\prime_1\mu_\chi v^\prime_\rho),
\nonumber \\ &&
{\bf M_{36}}=(\frac{g^2}{6}+g^{\prime2})v^\prime_\rho v_\chi+\frac{1}{12\sqrt2}
(-2f_1\mu_\rho v_\eta+\sqrt2 f_3\mu_\rho v_{\sigma_2}-2f^\prime_1\mu_\chi
v^\prime_\eta+\sqrt2f^\prime_3\mu_\chi v^\prime_{\sigma_2}),\nonumber \\ &&
{\bf M_{37}}=-(\frac{g^2}{3}+g^{\prime2})v_\chi v^\prime_\chi,\;
{\bf M_{38}}=-\frac{g^2}{12}v^\prime_{\sigma_2}v_\chi+\frac{1}{12}(f_3\mu_s
v_\rho+f^\prime_3\mu_\chi v^\prime_\rho),\nonumber \\ &&
{\bf M_{44}}=\frac{g^2}{12}v^2_{\sigma_2}+\frac{1}{18\sqrt2 v_{\sigma_2}}(f_1f_3v_\eta
v^2_\rho-\frac{18}{\sqrt2}k_3v_\rho v_\chi-\frac{3}{\sqrt2}f_3\mu_\rho
v^\prime_\rho v_\chi+f_1f_3 v_\eta
v^2_\rho+\frac{3}{\sqrt2}f^\prime_3\mu_s v^\prime_\rho
v^\prime_\chi\nonumber \\ &-&\frac{3}{\sqrt2}f_3\mu_\chi v_\rho
v^\prime_\chi),\; {\bf M_{45}}=\frac{g^2}{6}v^\prime_\eta v_{\sigma_2},\;\;
{\bf M_{46}}=-\frac{g^2}{12}v^\prime_\rho
v_{\sigma_2}+\frac{1}{12}(f_3\mu_\rho v_\chi+f^\prime_3\mu_s
v^\prime_\chi),\nonumber \\ &&
{\bf M_{47}}=-\frac{g^2}{12}v_{\sigma_2}v^\prime_\chi+
\frac{1}{12}(f^\prime_3\mu_s v^\prime_\rho+f_3\mu_\chi v_\rho),\;
{\bf M_{48}}=-\frac{g^2}{12}v_{\sigma_2}v^\prime_{\sigma_2},\nonumber \\ &&
{\bf M_{55}}=\frac{g^2}{3}v^{\prime2}_\eta+\frac{1}{18\sqrt2 v^\prime_\eta}
(f^\prime_1f^\prime_3v^{\prime2}_\rho v^\prime_{\sigma_2}-3f_1\mu_\rho v_\rho
v_\eta+3f^\prime_1\mu_\chi v^\prime_\rho v_\chi-18k^\prime_1 v^\prime_\rho
v^\prime_\chi+3f^\prime_1\mu_\rho v_\rho v^\prime_\chi\nonumber \\
&+&f^\prime_1f_3 v^\prime_{\sigma_2} v^{\prime2}_\chi),\;
{\bf M_{56}}=-\frac{g^2}{6}v^\prime_\eta
v^\prime_\rho+\frac{1}{9\sqrt2}(\sqrt2f^{\prime2}_1v^\prime_\eta
v^\prime_\rho-f^\prime_1f^\prime_3v^\prime_\rho v^\prime_{\sigma_2}-
\frac{3}{2}f^\prime_1\mu_\chi
v_\chi+9k^\prime_1 v^\prime_\chi),\nonumber \\ &&
{\bf M_{57}}=-\frac{g^2}{6}v^\prime_\eta v^\prime_\chi+
\frac{1}{9\sqrt2}(9k^\prime_1 v^\prime_\rho-\frac{3}{2}f^\prime_1\mu_\rho
v_\rho+\sqrt2 f^{\prime2}_1 v^\prime_\eta v^\prime_\chi-f^\prime_1f^\prime_3
v_{\sigma_2}v^\prime_\chi),\; {\bf M_{58}}=-\frac{g^2}{6}v^\prime_\eta
v^\prime_{\sigma_2}\nonumber \\
&-&\frac{f^\prime_1f^\prime_3}{18\sqrt2}(v^{\prime2}_\rho 
+v^{\prime2}_\chi),\;{\bf M_{66}}=(\frac{g^2}{3}+g^{\prime2})v^{\prime2}_\rho+
\frac{1}{12\sqrt2 v^\prime_\rho}(2f_1\mu_\rho v_\eta v_\chi-\sqrt2f_3\mu_\rho
v_{\sigma_2}v_\chi+2f^\prime_1\mu_\chi v^\prime_\eta v_\chi\nonumber \\
&-&\sqrt2 
f^\prime_3\mu_\chi v^\prime_{\sigma_2}v_\chi-12k^\prime_1v^\prime_\eta
v^\prime_\chi
-\frac{12}{\sqrt2}k^\prime_3v^\prime_{\sigma_2}v^\prime_\chi-2f^\prime_1\mu_\eta
v_\eta v^\prime_\chi-\sqrt2 f^\prime_3\mu_s v_{\sigma_2}v^\prime_\chi),\nonumber
\\ && {\bf M_{67}}=-(\frac{g^2}{6}+g^{\prime2})v^\prime_\rho
v^\prime_\chi+\frac{1}{12\sqrt2}(12k^\prime_1v^\prime_\eta+
\frac{12}{\sqrt2}k^\prime_3v^\prime_{\sigma_2}+2f^\prime_1\mu_\eta
v_\eta+\sqrt2f^\prime_3\mu_s
v_{\sigma_2}+\frac{3}{\sqrt2}f^{\prime2}_1v^\prime_\rho v^\prime_\chi\nonumber
\\&+&
\frac{3}{\sqrt2}f^{\prime2}_3v^\prime_\rho v^\prime_\chi),\;
{\bf M_{68}}=\frac{g^2}{12}v^\prime_\rho
v^\prime_{\sigma_2}+\frac{f^\prime_3v^\prime_\rho}{18\sqrt2}
(\sqrt2f^\prime_3v^\prime_{\sigma_2}-2f^\prime_1v^\prime_\eta),\nonumber \\ &&
{\bf M_{77}}=(\frac{g^2}{3}+g^{\prime2})v^{\prime2}_\chi-\frac{1}{12\sqrt2
v^\prime_\chi}(12k^\prime_1v^\prime_\eta v^\prime_\rho+12k^\prime_3v^\prime_\rho
v^\prime_{\sigma_2}+2f^\prime_1\mu_\eta v_\eta v^\prime_\rho-2f^\prime_1\mu_\rho
v^\prime_\eta v_\rho+\sqrt2f^\prime_3\mu_\rho v_\rho v_{\sigma_2}\nonumber \\
&+&\sqrt2f^\prime_3\mu_s v^\prime_\rho v_{\sigma_2}-2f_1\mu_\chi v_\eta
v_\rho+\sqrt2f_3\mu_\chi v_\rho v_{\sigma_2}),\;
{\bf M_{78}}=\frac{g^2}{12}v^\prime_{\sigma_2}v^\prime_\chi+\frac{1}{12}
(6k^\prime_3v^\prime_\rho+f^\prime_3\mu_\rho
v_\rho)\nonumber \\ &+&\frac{f^\prime_3 
v^\prime_\chi}{18\sqrt2}(f^\prime_3v^\prime_{\sigma_2}-f^\prime_1
v^\prime_\eta), \nonumber \\ &&
{\bf M_{88}}=\frac{g^2}{12}v^{\prime2}_{\sigma_2}-
\frac{1}{18\sqrt2v^\prime_{\sigma_2}}
(f^\prime_1f^\prime_3 v^\prime_\eta v^{\prime2}_\rho+\frac{3}{\sqrt2}
f_3\mu_s v_\rho v_\chi+\frac{3}{\sqrt2}f^\prime_3\mu_\chi
v^\prime_\rho v_\chi 
+\frac{18}{\sqrt2}k^\prime_3v_\rho
v^\prime_\chi+\frac{3}{\sqrt2}f^\prime_3\mu_\chi v^\prime_\rho v_\chi
\nonumber \\ &+&
\frac{18}{\sqrt2}k^\prime_3 v^\prime_\rho v^\prime_\chi+
\frac{3}{\sqrt2}f^\prime_3\mu_\rho v_\rho v^\prime_\chi-f^\prime_1f^\prime_3
v^\prime_\eta v^{\prime2}_\chi).
\label{mmr}
\end{eqnarray*}

\begin{figure}[ht]
\begin{center}
\vglue -0.009cm
\mbox{\epsfig{file=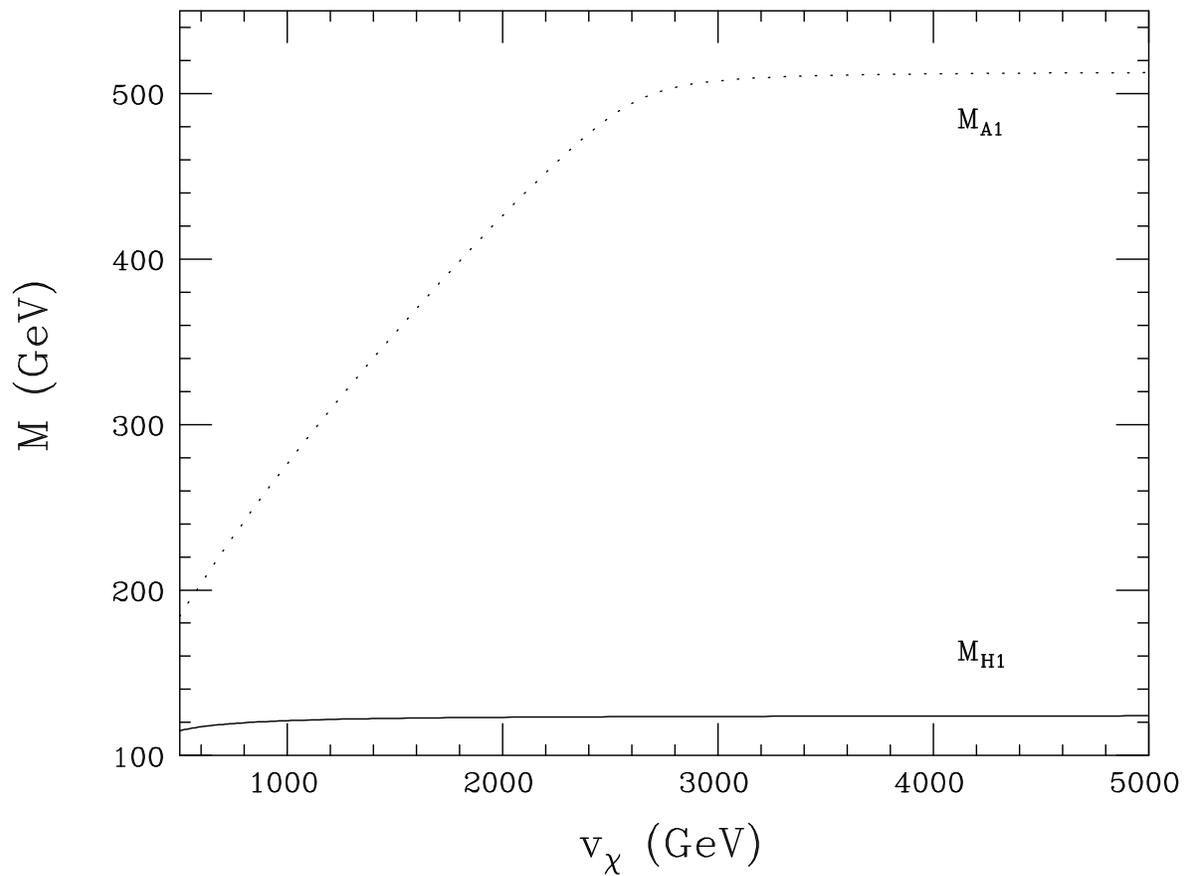,width=0.7\textwidth,angle=90}}       
\end{center}
\vglue 2cm
\caption{The eigenvalues of the mass matrix given in Appendix~\ref{sec:a1},
$M_{H_1}$ is the mass of the lightest scalar and $M_A$ the mass of the lightest
pseudoscalar.}
\label{fig1}
\end{figure}

This mass matrix has no Goldstone bosons and 8 mass eigenstates. Some typical
values of the masses of these scalars, for a set of values of the parameters,
are given in the text. In Fig.~\ref{fig1} we show the behavior of the mass of
the lightest scalar $H_1$ with the $v_\chi$, the largest VEV in the model. 

\section{Mass matrix of the pseudoscalar neutral fields}
\label{sec:a2}

The complete symmetric mass matrix in the CP-odd scalar sector, with the
constraint equation of the Appendix~\ref{sec:a3} taken into account, is given by
\begin{eqnarray*}
&&{\bf {\cal M}_{11}}=\frac{1}{18\sqrt2 v_\eta}(f_1f_3 v^2_\rho v_{\sigma_2}-18
k_1v_\rho v_\chi+3f_1\mu_\rho v^\prime_\rho v^\prime_\chi),\;
{\bf {\cal M}_{12}}=\frac{1}{6\sqrt2}(f_1\mu_\chi-6f_1 v_\chi),\nonumber \\ && 
{\bf {\cal M}_{13}}=\frac{1}{6\sqrt2}(f_1\mu_\rho v^\prime_\rho-6k_1v_\rho),\;
{\bf {\cal M}_{14}}=\frac{f_1f_3}{18\sqrt2}(v^2_\rho+v^2_\chi),\;
{\bf {\cal M}_{15}}=0, \nonumber \\ && 
{\bf {\cal M}_{16}}=\frac{1}{6\sqrt2}(f_1\mu_\eta
v^\prime_\chi-f^\prime_1\mu_\rho v_\chi),\;{\bf {\cal M}_{17}}=
\frac{1}{6\sqrt2}(f^\prime_1\mu_\eta-f_1\mu_\chi 
v_\rho),\;{\bf {\cal M}_{18}}=0, \nonumber \\ && 
{\bf {\cal M}_{22}}=\frac{1}{6\sqrt2 v_\rho}(-6k_1 v_\eta
v_\chi-\frac{6}{\sqrt2}k_3v_{\sigma_2}v_\chi-f_1\mu_\eta v^\prime_\eta v_\chi-
\frac{1}{\sqrt2}f_3\mu_s v^\prime_{\sigma_2}v_\chi+f^\prime_1\mu_\rho
v^\prime_\eta v^\prime_\chi\nonumber \\&-&\frac{1}{\sqrt2}f^\prime_3\mu_\rho
v^\prime_{\sigma_2}v^\prime_\chi+f_1\mu_\chi v_\eta
v^\prime_\chi-\frac{1}{\sqrt2}f_3\mu_\chi v_{\sigma_2}v^\prime_\chi),\;
{\bf {\cal M}_{23}}=\frac{1}{6\sqrt2}(-6k_1v_\eta-
\frac{6}{\sqrt2}k_3v_{\sigma_2}-f_1\mu_\eta v^\prime_\eta \nonumber \\ &-&
\frac{1}{\sqrt2}\mu_s v^\prime_{\sigma_2}),\;
{\bf {\cal M}_{24}}=\frac{1}{12}(6k_3
v_\chi+f_3\mu_\chi v^\prime_\chi), \;{\bf {\cal M}_{25}}=
\frac{1}{6\sqrt2}(f_1\mu_\eta v_\chi
-f^\prime_1\mu_\rho v^\prime_\chi),\;{\cal M}_{26}=0,\nonumber \\ &&
{\bf {\cal M}_{27}}=\frac{1}{12\sqrt2}(-2f^\prime_1\mu_\rho v^\prime_\eta+\sqrt2
f^\prime_3\mu_\rho v^\prime_{\sigma_2}-2f_1\mu_\chi v_\eta+\sqrt2 f_3\mu_\chi
v_{\sigma_2}), \nonumber \\ &&{\bf {\cal M}_{28}}=\frac{1}{12}(f_3\mu_s
v_\chi+f^\prime_3\mu_\rho v^\prime_\chi),\;
{\bf {\cal M}_{33}}=\frac{1}{6\sqrt2 v_\chi}(-6f_1 k_1v_\eta
v_\rho-\frac{6}{\sqrt2}k_3 v_\rho v_{\sigma_2}-f_1\mu_\eta v^\prime_\eta
v_\rho+f_1\mu_\rho v_\eta 
v^\prime_\rho \nonumber \\ &-&\frac{1}{\sqrt2}f_3\mu_\rho v^\prime_\rho
v_{\sigma_2}-\frac{1}{\sqrt2}f_3\mu_s v_\rho v^\prime_{\sigma_2}+
f^\prime_1\mu_\chi v^\prime_\eta v^\prime_\rho-\frac{1}{\sqrt2}f^\prime_3
\mu_\chi v^\prime_\rho v^\prime_{\sigma_2}),\;{\bf {\cal M}_{34}}=
\frac{1}{12}(6k_3v_\rho+f_3 \mu_\rho v^\prime_\rho),\nonumber \\ &&
{\bf {\cal M}_{35}}=\frac{1}{\sqrt2}(f_1\mu_\eta v_\rho-f^\prime_1\mu_\chi
v^\prime_\chi),\; {\bf {\cal M}_{36}}=
\frac{1}{12\sqrt2}(-2f_1\mu_\rho v_\eta+\sqrt2
f_3\mu_\rho v_{\sigma_2}-2f^\prime_1\mu_\chi v^\prime_\eta\nonumber \\ &+&\sqrt2
f^\prime_3\mu_\chi v^\prime_{\sigma_2}),\;{\bf {\cal M}_{37}}=0,\;
{\bf {\cal M}_{38}}=\frac{1}{12}(f_3\mu_s v_\rho+f^\prime_3\mu_\chi
v^\prime_\rho), \nonumber \\ && {\bf {\cal M}_{44}}=\frac{1}{18\sqrt2
v_{\sigma_2}}(f_1f_3 v_\eta v^2_\rho-\frac{18}{\sqrt2}k_3 v_\rho
v_\chi-\frac{3}{\sqrt2}\mu_\rho v^\prime_\rho v_\chi+\frac{18}{\sqrt2}k_3v_\rho
v_\chi-\frac{3}{\sqrt2}f_3\mu_\rho v^\prime_\rho v_\chi
\nonumber \\ &+&f_1f_3 v_\eta v^2_\chi
-\frac{3}{\sqrt2}f^\prime_3\mu_s v^\prime_\rho v^\prime_\chi-
\frac{3}{\sqrt2}f_3\mu_\chi v_\rho v^\prime_\chi),\; {\cal M}_{45}=0,\;
{\bf {\cal M}_{46}}=-\frac{1}{12}(f_3\mu_\rho v_\chi+f^\prime_3\mu_s
v^\prime_\chi), \nonumber \\ && {\bf {\cal M}_{47}}=
-\frac{1}{12}(f^\prime_3\mu_\rho v^\prime_\rho 
+f_3\mu_\chi v^\prime_\rho),\; {\bf {\cal M}_{48}}=0,\nonumber \\ &&
{\bf {\cal M}_{55}}=\frac{1}{18\sqrt2 v^\prime_\eta}(f^\prime_1f^\prime_3
v^{\prime2}_\rho v^\prime_{\sigma_2}-3f_1\mu_\eta v_\rho v_\chi+3f^\prime_1
\mu_\chi v^\prime_\rho v_\chi-18 k^\prime_1 v^\prime_\rho
v^\prime_\chi+3f^\prime_1\mu_\rho v_\rho v^\prime_\chi+f^\prime_1f^\prime_3
v^\prime_{\sigma_2}v^{\prime2}_\chi\nonumber \\ &+&
3f^\prime_1\mu_\rho v_\rho v^\prime_\chi
+f^\prime_1f^\prime_3 v^\prime_{\sigma_2} v^{\prime2}_\chi),\;
{\bf {\cal M}_{56}}=\frac{f_1f_3}{18\sqrt2}(\mu_\chi v_\chi-k^\prime_1
v^\prime_\chi),\;\; {\bf {\cal M}_{57}}=\frac{1}{6\sqrt2}(-6k^\prime_1 
v^\prime_\rho+f^\prime_1\mu_\rho v_\rho),\nonumber \\ &&
{\bf {\cal M}_{58}}=\frac{f^\prime_1f^\prime_3}{18\sqrt2}(v^{\prime2}-v^{\prime2}),
\; {\bf {\cal M}_{66}}=\frac{1}{12\sqrt2 v^\prime_\rho}(2f_1\mu_\rho v_\eta v_\chi
-\sqrt2 f_3\mu_\rho v_{\sigma_2}v_\chi+2f^\prime_1\mu_\chi v^\prime_\eta v_\chi
\nonumber \\ &-&
\sqrt2 f^\prime_3\mu_\chi v^\prime_{\sigma_2}v_\chi-12k^\prime_1 v^\prime_\eta
v^\prime_\chi
-\frac{12}{\sqrt2}k^\prime_3 v^\prime_{\sigma_2}
v^\prime_\chi-2f^\prime_1\mu_\eta v_\eta v^\prime_\chi-\sqrt2 \mu_s v_{\sigma_2}
v^\prime_\chi),\nonumber \\ && {\cal M}_{67}=\frac{1}{12\sqrt2}(-12k^\prime_1
v^\prime_\eta-\frac{12}{\sqrt2}k^\prime_3
v^\prime_{\sigma_2}-2f^\prime_1\mu_\eta v_\eta-\sqrt2 f^\prime_3\mu_s
v_{\sigma_2}),\; {\bf {\cal M}_{68}}=\frac{1}{12}(-f^\prime_3\mu_\chi
v_\chi+6k^\prime_3 v^\prime_\chi),\nonumber \\ &&
{\bf {\cal M}_{77}}=\frac{1}{12\sqrt2 v^\prime_\chi}(-12k^\prime_3 v^\prime_\eta
v^\prime_\rho-\frac{12}{\sqrt2}k^\prime_3 v^\prime_\rho
v^\prime_{\sigma_2}-2f^\prime_1\mu_\eta v_\eta v^\prime_\rho+
2f^\prime_1\mu_\rho v^\prime_\eta v_\rho-\sqrt2f^\prime_3\mu_\rho v_\rho
v^\prime_{\sigma_2}), \nonumber \\ &&
{\bf {\cal M}_{78}}=\frac{1}{12}(6k^\prime_3 v^\prime_\rho-f^\prime_3 \mu_\rho
v_\rho),\;{\bf {\cal M}_{88}}=\frac{1}{18\sqrt2
v^\prime_{\sigma_2}}(f^\prime_1f^\prime_3 v^\prime_\eta v^{\prime2}_\rho-
\frac{3}{\sqrt2}\mu_s v_\rho v_\chi-\frac{3}{\sqrt2}\mu_\chi v^\prime_\rho
v_\chi-\frac{18}{\sqrt2}k^\prime_3v^\prime_\rho v^\prime_\chi
\nonumber \\ &-&
\frac{3}{\sqrt2}f^\prime_3\mu_\rho v_\rho v^\prime_\chi+f^\prime_1f^\prime_3
v^\prime_\eta v^{\prime2}).
\label{mmp}
\end{eqnarray*}

This mass matrix has two Goldstone bosons and 6 mass eigenstates. In
Fig.~\ref{fig1} we show the behavior of the mass of the lightest pseudoscalar 
scalar $A_1$ as a function of $v_\chi$. Typical values for the masses in this
sector for a set of values of the parameters are given in the text.

\section{Constraints equations}
\label{sec:a3}

The constraints equations are

\begin{eqnarray*}
&&{\bf \frac{t_\eta}{v_\eta}}=
\frac{g^2}{12}(2v^2_\eta-2v^{\prime2}_\eta-v^2_\rho+v^{\prime2}_\rho-
v^2_{\sigma_2}+v^{\prime2}_{\sigma_2}-v^2_\chi+v^{\prime2}_\chi)
+m^2_\eta+\frac{1}{4}\mu^2_\eta+\frac{f^2_1}{18}(v^2_\rho+v^2_\chi)
\nonumber \\ &+&\frac{f_1f_3}{18\sqrt2}\frac{v_{\sigma_2}}{v_\eta}(v^2_\rho
+v^2_\chi)+\frac{k_1}{\sqrt2}\frac{v_\rho v_\chi}{v_\eta}+
\frac{1}{6\sqrt2 v_\eta}(f_1\mu_\rho v^\prime_\rho v_\chi+f_1\mu_\chi v_\rho
v^\prime_\chi+f^\prime_1\mu_\eta v^\prime_\rho v^\prime_\chi),
 \\ &&
{\bf \frac{t_\rho}{v_\rho}}=
\frac{g^2}{12}(-v^2_\eta+v^{\prime2}_\eta+2v^2_\rho-
2v^{\prime2}+\frac{1}{2}v^2_{\sigma_2}-\frac{1}{2}v^{\prime2}_{\sigma_2}-
v^2_\chi+v^{\prime2}_\chi)+\frac{g^{\prime2}}{2}(v^2_\rho-v^{\prime2}_\rho-
v^2_\chi+v^{\prime2}_\chi)\nonumber \\ &+&
\frac{f^2_1}{18}(v^2_\eta+v^2_\chi)+\frac{f^2_3}{36}(v^2_{\sigma_2}+2v^2_\chi)
-\frac{f_1f_3}{9\sqrt2}v_\eta v_{\sigma_2}+m^2_\rho+\frac{1}{4}\mu^2_\rho+
\frac{k_1}{\sqrt2}\frac{v_\eta v_\chi}{v_\rho}+
\frac{k_3}{2}\frac{v_{\sigma_2}v_\chi}{v_\rho} \nonumber \\ &+&
\frac{f_1}{6\sqrt2 v_\rho}(\mu_\eta v^\prime_\eta v_\chi+\mu_\chi v_\eta
v^\prime_\chi)+\frac{f_3}{12v_\rho}(\mu_s v^\prime_{\sigma_2}v_\chi+\mu_\chi
v_{\sigma_2}v^\prime_\chi)+\frac{f^\prime_3}{12v_\rho}\mu_\rho
v^\prime_{\sigma_2} v^\prime_\chi-\frac{f^\prime_1}{6\sqrt2 v_\rho}\mu_\rho
v^\prime_\eta v^\prime_\chi
\\ &&
{\bf \frac{t_\chi}{v_\chi}}=\frac{g^2}{12}(-v^2_\eta+v^{\prime2}_\eta-v^2_\rho+
v^{\prime2}_\rho+v^2_{\sigma_2}-v^{\prime2}_{\sigma_2}+2v^2_\chi-
2v^{\prime2}_\chi)+\frac{g^{\prime2}}{2}(v^2_\chi-v^{\prime2}_\chi-v^2_\rho
+v^{\prime2}_\rho)\nonumber \\ &+&
\frac{f^2_1}{18}(v^2_\eta+v^2_\eta)+\frac{f^2_3}{36}(2v^2_\rho+v^2_{\sigma_2})
-\frac{f_1f_3}{9\sqrt2}v_\eta v_{\sigma_2}+m^2_\chi+\frac{1}{4}\mu^2_\chi
-\frac{1}{6\sqrt2 v_\chi}(f_1\mu_\rho v_\eta v^\prime_\rho-f_1\mu_\eta
v^\prime_\eta v_\rho \nonumber \\ &+&f^\prime_1\mu_\chi v^\prime_\eta v^\prime_\rho)
+\frac{1}{12 v_\chi}(f_3\mu_\rho v^\prime_\rho
v_{\sigma_2}+f_3\mu_s v_\rho 
v_{\sigma_2}+f^\prime_3\mu_\chi v^\prime_\rho v^\prime_{\sigma_2})
+\frac{k_1}{\sqrt2}\frac{v_\eta v_\rho}{v_\chi}+\frac{k_3}{2}\frac{v_\rho
v_{\sigma_2}}{v_\chi},
\\ &&
{\bf \frac{t_{\sigma_2}}{v_{\sigma_2}}}=
\frac{g^2}{24}(-2v^2_\eta+2v^{\prime2}_\eta
+v^2_\rho-v^{\prime2}_\rho-v^{\prime2}_{\sigma_2}+v^2_{\sigma_2}+ v^2_\chi
-v^{\prime2}_\chi)+\frac{f^2_3}{36}(v^2_\rho+v^2_\chi)+m^2_s
+\frac{1}{4}\mu_s^2\nonumber
\\ &-& \frac{f_1f_3v_\eta}{18\sqrt2v_{\sigma_2}}(v^2_\rho+v^2_\chi)
+\frac{1}{12v_{\sigma_2}}(f^\prime_3\mu_s v^\prime_\rho
v^\prime_\chi+f_3\mu_\chi v_\rho v^\prime_\chi+f_3\mu_\rho v^\prime_\rho v_\chi)
+\frac{k_3}{2}\frac{v_\rho v_\chi}{v_{\sigma_2}},
\\ &&
{\bf
\frac{t_{\eta^\prime}}{v^\prime_\eta}}=
\frac{g^2}{12}(-2v^2_\eta+v^{\prime2}_\eta
+v^2_\rho-v^{\prime2}_\rho+v^2_{\sigma_2}-v^{\prime2}_{\sigma_2}+v^2_\chi
-v^{\prime2}_\chi)+m^2_{\eta^\prime}+\frac{1}{4}\mu^2_\eta+
\frac{f^2_1}{18}(v^{\prime2}_\rho+
v^{\prime2}_\chi)\nonumber \\ &-&\frac{f_1f_3}{18\sqrt2}
\frac{v^\prime_{\sigma_2}}{v^\prime_\eta}(v^{\prime2}_\rho+v^{\prime2}_\chi)
+\frac{k^\prime_1}{\sqrt2}\frac{v^\prime_\rho v^\prime_\chi}{v^\prime_\eta}
+\frac{1}{6\sqrt2 v^\prime_\eta}(f_1\mu_\eta v_\rho v_\chi-f^\prime_1\mu_\chi
v^\prime_\rho v_\chi-
f^\prime_2\mu_\rho v_\rho v^\prime_\chi),
\\ && 
{\bf \frac{t_{\rho^\prime}}{v^\prime_\rho}}=
\frac{g^2}{12}(v^2_\eta-v^{\prime2}_\eta-2v^2_\rho+2v^{\prime2}_\rho-
\frac{1}{2}v^2_{\sigma_2}+\frac{1}{2}v^{\prime2}_{\sigma_2}+v^2_\chi-
v^{\prime2}_\chi)+\frac{g^{\prime2}}{2}(-v^2_\rho+v^{\prime2}_\rho+v^2_\chi
-v^{\prime2}_\chi)\nonumber \\ &+&
\frac{f^\prime_1}{18}(v^{\prime2}_\eta+v^{\prime2}_\chi)+
\frac{f^{\prime2}_3}{36}(v^{\prime2}_{\sigma_2}+2v^{\prime2}_\chi)-
\frac{f^\prime_1f^\prime_3}{9\sqrt2}v^\prime_\eta v^\prime_{\sigma_2}+
m^2_{\rho^\prime}+\frac{1}{4}\mu^2_\rho+
\frac{1}{6\sqrt2 v^\prime_\rho}(f_1\mu_\rho v_\eta v_\chi-f^\prime_1\mu_\chi
v^\prime_\eta v_\chi\nonumber \\ &+&f^\prime_1\mu_\eta v_\eta v^\prime_\chi)
+\frac{1}{12v^\prime_\rho}(f_3\mu_\rho v_{\sigma_2}v_\chi+f^\prime_3\mu_\chi
v^\prime_{\sigma_2}v_\chi+f^\prime_3\mu_s v_{\sigma_2}v^\prime_\chi)
+\frac{k^\prime_1}{\sqrt2}
\frac{v^\prime_\eta v^\prime_\chi}{v^\prime_\rho}+\frac{k^\prime_3}{3}
\frac{v^\prime_{\sigma_2}v^\prime_\chi }{v^\prime_\rho}, \\ &&
{\bf \frac{t_{\chi^\prime}}{v^\prime_\chi}}=
\frac{g^2}{12}(v^2_\eta-v^{\prime2}_\eta+v^2_\rho-v^{\prime2}_\rho+
\frac{1}{2}v^{\prime2}_{\sigma_2}-\frac{1}{2}v^2_{\sigma_2}-v^2_\chi
+v^{\prime2}_\chi)+\frac{g^{\prime2}}{2}(v^2_\rho-v^{\prime2}_\rho -v^2_\chi
+v^{\prime2}_\chi)\nonumber \\ &+&
\frac{f^{\prime2}_1}{18}(v^{\prime2}_\eta+v^2_\rho)+
\frac{f^{\prime2}_3}{36}(2v^{\prime2}_\rho+v^{\prime2}_{\sigma_2})+
\frac{f^\prime_1f^\prime_3}{9\sqrt2}v^\prime
v^\prime_{\sigma_2}+m^2_{\chi^\prime}+\frac{1}{4}\mu^2_\chi+
\frac{1}{6\sqrt2 v^\prime_\chi}(f^\prime_1\mu_\eta v_\eta v^\prime_\rho-
f^\prime_1\mu_\rho v^\prime_\eta v_\rho \nonumber \\ &+&
f_1\mu_\chi v_\eta v_\rho)
+\frac{1}{12v^\prime_\chi}(f^\prime_3 \mu_\rho v_\rho v^\prime_{\sigma_2}+
f^\prime_3\mu_s v^\prime_\rho v_{\sigma_2}+f_3\mu_\chi v_\rho v_{\sigma_2})+
\frac{k^\prime_1}{\sqrt2}\frac{v^\prime_\eta v^\prime_\rho}{v^\prime_\chi}+
\frac{k^\prime_3}{2}\frac{v^\prime_\rho v^\prime_{\sigma_2}}{v^\prime_\chi},
\\ &&
{\bf \frac{t_{\sigma^\prime_2}}{v^\prime_{\sigma_2}}}=
\frac{g^2}{24}(2v^2_\eta-2v^{\prime2}_\eta-v^2_\rho+v^{\prime2}_\rho-
v^2_{\sigma_2}+v^{\prime2}_{\sigma_2}-v^2_\chi+v^{\prime2}_\chi)+
\frac{f^{\prime2}_3}{36}(v^{\prime2}_\rho+v^{\prime2}_\chi)+
m^2_{\sigma^\prime_2}+\frac{1}{4}\mu^2_s\nonumber \\ &-&
\frac{f^\prime_1f^\prime_3}{18\sqrt2}\frac{v^\prime_\eta}{ v^\prime_{\sigma_2}}
(v^\prime_\rho+v^{\prime2}_\chi)+
\frac{1}{12v^\prime_{\sigma_2}}(f_3\mu_s v_\rho v_\chi+f^\prime_3\mu_\chi
v^\prime_\rho v_\chi+f^\prime_3 \mu_\rho v_\rho v^\prime_\chi)
+\frac{k^\prime_3}{2}\frac{v^\prime_\rho v^\prime_\chi}{v^\prime_{\sigma_2}}.
\end{eqnarray*}

\vglue 0.01cm
\end{document}